\documentclass[10pt]{iopart}

\usepackage{graphicx}
\usepackage{times}

\begin{document}

\title{Uncertainties of Euclidean Time Extrapolation in Lattice Effective Field Theory}

\author{Timo A. L\"ahde$^1$, Evgeny Epelbaum$^2$, Hermann Krebs$^2$, 
Dean Lee$^3$, Ulf-G.~Mei{\ss}ner$^{1,4,5}$, and Gautam Rupak$^6$} 

\address{$^1$Institute~for~Advanced~Simulation, Institut~f\"{u}r~Kernphysik, and 
J\"{u}lich Center for Hadron Physics, Forschungszentrum~J\"{u}lich, D--52425~J\"{u}lich, Germany}
\address{$^2$Institut~f\"{u}r~Theoretische~Physik~II,~Ruhr-Universit\"{a}t~Bochum, D--44870~Bochum,~Germany}
\address{$^3$Department~of~Physics, North~Carolina~State~University, Raleigh, NC~27695, USA}
\address{$^4$Helmholtz-Institut f\"ur Strahlen- und Kernphysik and Bethe Center for Theoretical Physics,
Universit\"at Bonn, D--53115 Bonn, Germany}
\address{$^5$JARA~--~High~Performance~Computing, Forschungszentrum~J\"{u}lich, D--52425 J\"{u}lich, Germany}
\address{$^6$Department~of~Physics~and~Astronomy, Mississippi~State~University, Mississippi State, MS~39762, USA}

\eads{\mailto{t.laehde@fz-juelich.de}, \mailto{evgeny.epelbaum@rub.de}, \mailto{hermann.krebs@rub.de}, 
\mailto{dean\_lee@ncsu.edu}, \mailto{u.meissner@fz-juelich.de}, \mailto{grupak@u.washington.edu}}

\begin{abstract}
Extrapolations in Euclidean time form a central part of Nuclear Lattice Effective Field Theory (NLEFT) calculations using
the Projection Monte Carlo method, as the sign problem in many cases prevents
simulations at large Euclidean time. We review the next-to-next-to-leading order NLEFT results for the alpha nuclei up to $^{28}$Si, 
with emphasis on the Euclidean time extrapolations, their expected accuracy and potential pitfalls. We also discuss possible avenues for 
improving the reliability of Euclidean time extrapolations in NLEFT.
\end{abstract}


\vspace{2pc}
\maketitle


\section{Introduction}

Several \textit{ab initio} methods are currently being applied to the study of nuclear structure. These include
coupled-cluster expansions~\cite{Hagen:2012fb}, the no-core shell model~\cite{Jurgenson:2013yya,Roth:2011ar}, the 
in-medium similarity renormalization group~\cite{Hergert:2012nb}, self-consistent Green's functions~\cite{Soma:2012zd}, 
Green's function Monte Carlo~\cite{Lovato:2013cua}, and Auxiliary-Field Diffusion Monte Carlo \cite{Gandolfi:2014ewa}. 

Much of the recent progress in \textit{ab initio} nuclear structure calculations is due to
``soft'' chiral nuclear Effective Field Theory~(EFT) interactions. 
The lattice formulation of chiral nuclear EFT is described in Ref.~\cite{Borasoy:2006qn}, and a review of lattice EFT methods can be found in
Ref.~\cite{Dean_QMC}. A comprehensive overview of chiral nuclear EFT is available in Refs.~\cite{Epelbaum:2008ga,Machleidt:2011zz}.
This framework, known as Nuclear Lattice Effective Field Theory (NLEFT), has been used to calculate the ground states of alpha nuclei 
from $^{4}$He to $^{28}$Si, as well as to describe the structure of the Hoyle state~\cite{Epelbaum:2011md,Epelbaum:2012qn} 
and the dependence of the triple-alpha process on the fundamental parameters of nature~\cite{Epelbaum:2012iu}.

NLEFT is an \textit{ab initio} method where chiral nuclear EFT is combined with Auxiliary-Field Quantum Monte Carlo (AFQMC) lattice calculations. 
This Monte Carlo approach differs from other \textit{ab initio} methods in that it does not require truncated basis 
expansions, many-body perturbation theory, or any constraint on the nuclear wave function. While our NLEFT results are thus unbiased Monte Carlo 
calculations, the AFQMC approach nevertheless relies on Euclidean time projection of Projection Monte Carlo (PMC) data in order to compute the properties of the 
low-lying states of light and medium-mass nuclei. One of the largest sources of computational uncertainty is then due to the appearance of complex 
sign oscillations or the ``sign problem'' which limits the extent of Euclidean time available for direct PMC calculations. While the sign problem is greatly
suppressed by the soft interaction employed in NLEFT, it still represents a significant obstacle to practical PMC calculations, especially in cases
where the number of protons is not equal to the number of neutrons.

In this paper, we focus on the question of the accuracy and reliability of the Euclidean time extrapolations. We start in Section~\ref{PMC} by reviewing the PMC formalism,
along with the methodology for extrapolating to infinite Euclidean time in Section~\ref{extra}. Next, we provide in Section~\ref{MC} an overview of the NLEFT results for the
alpha nuclei ranging from $^{4}$He to $^{28}$Si, as obtained from Euclidean time extrapolations of PMC data corresponding to the  
lattice action described in Refs.~\cite{Epelbaum:2011md,Epelbaum:2012qn,Epelbaum:2009zsa}. In Section~\ref{errors}, we consider the effects of
statistical and systematical errors on the accuracy of the extrapolation method, and we conclude in Section~\ref{outlook} by a discussion of future
improvements and refinements to our extrapolation methods.


\section{Projection Monte Carlo formalism \label{PMC}}

Our NLEFT calculations are, as in chiral nuclear EFT, organized in powers of a soft scale $Q$, which is associated with factors of momenta and the 
pion mass. The contributions of $\mathcal{O}(Q^0)$ to the nuclear Hamiltonian are referred to as leading order (LO), the $\mathcal{O}(Q^2)$ terms are of next-to-leading 
order (NLO), and the $\mathcal{O}(Q^3)$ terms are of next-to-next-to-leading order (NNLO), at which point our present calculations are truncated. The LO lattice
Hamiltonian includes a significant part of the NLO and higher-order contributions, as smeared contact interactions~\cite{Borasoy:2006qn,Epelbaum:2009pd,Epelbaum:2010xt} are
used. It should also be noted that since we are using a low-momentum power counting scheme, no additional two-nucleon contributions arise at NNLO beyond the 
terms already appearing at NLO, as these can be absorbed into redefinitions of the NLO couplings. A full discussion of the interactions used in the reported results
can be found in Ref.~\cite{Epelbaum:2009zsa}.

In the present NLEFT calculations, we have used a periodic $L = 6$ cube and a lattice spacing of $a = 1.97$~fm, which translates into a cube length of 
$La = 11.82$~fm. Our initial wave functions, $|\Psi_{A}^\mathrm{init}\rangle$, are Slater-determinant states composed of delocalized standing waves in the periodic cube
with $A$ nucleons. Localized alpha-cluster trial states have also been used for studies of $^{12}$C and $^{16}$O~\cite{Epelbaum:2011md,Epelbaum:2012qn,16O_spectrum}. 
These provide not only a consistency check on the Euclidean time extrapolation, but also an opportunity to assess the spatial structure of the nuclei. For simplicity, we describe 
our calculations using the language of continuous time evolution, even though our AFQMC calculations use transfer matrices with a temporal 
lattice spacing of $a_t^{}=1.32$~fm~\cite{Dean_QMC}. The Euclidean projection time is given by $t = N_t^{}a_t^{}$, where $N_t^{}$ denotes the number of Euclidean time slices.

We start the Euclidean time projection by means of a ``low-energy filter'' based upon Wigner's SU(4) symmetry, where 
the spin and isospin degrees of freedom of the nucleon are all equivalent as four components of an SU(4) multiplet. The SU(4) symmetric Hamiltonian is of the form
\begin{equation}
H_\mathrm{SU(4)}^{} \equiv H_\mathrm{free}^{} 
+\frac{1}{2} \, C_\mathrm{SU(4)}^{}\sum_{\vec n,\vec n'} 
{:\rho(\vec n)f(\vec n - \vec n')\,\rho(\vec{n}'):},
\label{H_SU4}
\end{equation}
where $f(\vec n - \vec n')$ is a Gaussian smearing function with its width set by the average effective range of the two $S$-wave interaction channels, 
and $\rho$ is the total nucleon density. Application of the exponential of $H_\mathrm{SU(4)}$ gives
\begin{equation}
|\Psi_A^{}(t^\prime_{})\rangle \equiv \exp(-H_\mathrm{SU(4)}^{} t^\prime_{}) |\Psi_{A}^\mathrm{init}\rangle,
\label{trial}
\end{equation}
referred to as a ``trial state''.  This part of the calculation is computationally straightforward, as it only requires a single 
auxiliary field. Most significantly, it does not generate any sign oscillations in the Monte Carlo calculation. 

Next, we use the
full LO Hamiltonian $H_\mathrm{LO}^{}$ to evolve the trial state for a time $t$, and 
construct the Euclidean-time projection amplitude
\begin{equation}
Z_A^{}(t) \equiv \langle\Psi_A^{}(t^\prime_{})| \exp(-H_\mathrm{LO}^{} t) |\Psi_A^{}(t^\prime_{})\rangle,
\end{equation}
from which we compute the ``transient energy''
\begin{equation}
E_A^{}(t) = -\partial[\ln Z_A^{}(t)]/\partial t,
\label{EAt}
\end{equation}
by means of a numerical finite difference. Hence, if
the lowest eigenstate of $H_\mathrm{LO}^{}$ that possesses a non-vanishing overlap with 
the trial state $|\Psi_A^{}(t^\prime_{})\rangle$ is denoted $|\Psi_{A,0}^{}\rangle$, the energy $E_{A,0}^{}$ of $|\Psi_{A,0}^{}\rangle$ 
is obtained as the ${t\to\infty}$ limit of $E_A^{}(t)$. Sign oscillations in the Monte Carlo calculation set the main limitation on the
number of Euclidean time steps for which Eq.~(\ref{EAt}) can be evaluated. The coupling
$C_\mathrm{SU(4)}^{}$ is a free parameter which can be used either to optimize the convergence of the Euclidean time evolution,
or to provide additional constraints for the extrapolation $t \to \infty$.

Higher-order corrections to $E_{A,0}^{}$ are evaluated using perturbation theory. We compute
expectation values using
\begin{equation}
Z_A^\mathcal{O}(t) \equiv
\langle\Psi_A^{}(t^\prime_{})| \exp(-H_\mathrm{LO}^{} t/2)
\mathcal{O} \exp(-H_\mathrm{LO}^{} t/2)  |\Psi_A^{}(t^\prime_{})\rangle,
\label{OP}
\end{equation}
for any operator $\mathcal{O}$. Given the ratio
\begin{equation}
X_A^\mathcal{O}(t) = Z_A^\mathcal{O}(t)/Z_A^{}(t),
\label{XAt}
\end{equation} 
the expectation
value of $\mathcal{O}$ for the desired state $|\Psi_{A,0}^{}\rangle$ is again obtained in the 
$t\rightarrow\infty$ limit according to
\begin{equation}
X_{A,0}^\mathcal{O} \equiv \langle\Psi_{A,0}^{}| \: \mathcal{O} \: |\Psi_{A,0}^{}\rangle = \lim_{t \to \infty}X_A^\mathcal{O}(t),
\end{equation}
which gives the corrections to $E_{A,0}^{}$ induced by the NLO and NNLO contributions, including the effects of strong and electromagnetic
isospin symmetry breaking.

The closer the trial state $|\Psi_A^{}(t^\prime_{})\rangle$ is to $|\Psi_{A,0}^{}\rangle$, 
the less the required projection time $t$. The trial state can be optimized by adjusting both the SU(4) projection time $t^\prime_{}$ and the strength 
of the coupling $C_\mathrm{SU(4)}^{}$ of $H_\mathrm{SU(4)}$. As shown in Section~\ref{MC}, the accuracy and reliability of the extrapolation
$t\to\infty$ is greatly improved by simultaneously incorporating data from multiple trial states 
that differ in the choice of $C_\mathrm{SU(4)}^{}$. This approach enables a ``triangulation'' of the asymptotic behavior as 
the common limit of several different functions of $t$. 


\section{Extrapolation in Euclidean time \label{extra}}

In most cases, reaching the limit $t \to \infty$ requires an extrapolation from finite values of $t$. Given the limited extent of the data, there are uncertainties in this extrapolation.
The behavior of $Z_A^{}(t)$ and $Z_A^{\mathcal{O}}(t)$ at large $t$ is controlled by the low-energy spectrum of $H_\mathrm{LO}^{}$. 
Let $| E\rangle$ label the eigenstates of $H_\mathrm{LO}^{}$ with energy $E$, and let $\rho_{A}^{}(E)$ denote the density of states for a system
of $A$~nucleons. For simplicity, we omit additional labels needed to distinguish degenerate states.  
We can then express $Z_A^{}(t)$ and $Z_A^{\mathcal{O}}(t)$ in terms of their spectral representations,
\begin{eqnarray}
Z_A^{}(t) & = & \int dE \: \rho_A^{}(E) \:
\big| \langle E |\Psi_A^{}(t^\prime_{})\rangle\big|^2_{} 
\exp(-Et), \\
Z_A^{\mathcal{O}}(t) & = & \int dE\,dE^\prime_{} \, \rho_A^{}(E)\,\rho_A^{}(E^\prime_{}) 
\, \exp(-(E+E^\prime_{})t/2), \nonumber \\ 
& \times &
\langle\Psi_A^{}(t^\prime_{})|E\rangle \,
\langle E|\mathcal{O}|E^\prime_{}\rangle \,
\langle E^\prime_{}|\Psi_A^{}(t^\prime_{})\rangle.
\label{reps}
\end{eqnarray} 
The spectral representations of $E_A^{}(t)$ and $X_A^{\mathcal{O}}(t)$ are then obtained by using Eq.~(\ref{EAt}) and Eq.~(\ref{XAt}), respectively. 
We can approximate these to arbitrary accuracy over any finite range of $t$ by taking $\rho_{A}^{}(E)$ to be a sum of energy delta functions,
\begin{equation}
\rho_{A}^{}(E) \approx \sum_{k=0}^{k_\mathrm{max}}c_{A,k}\delta(E-E_{A,k}^{}).
\label{delta}
\end{equation}
%


\begin{figure}[t]
\includegraphics[width=\columnwidth]{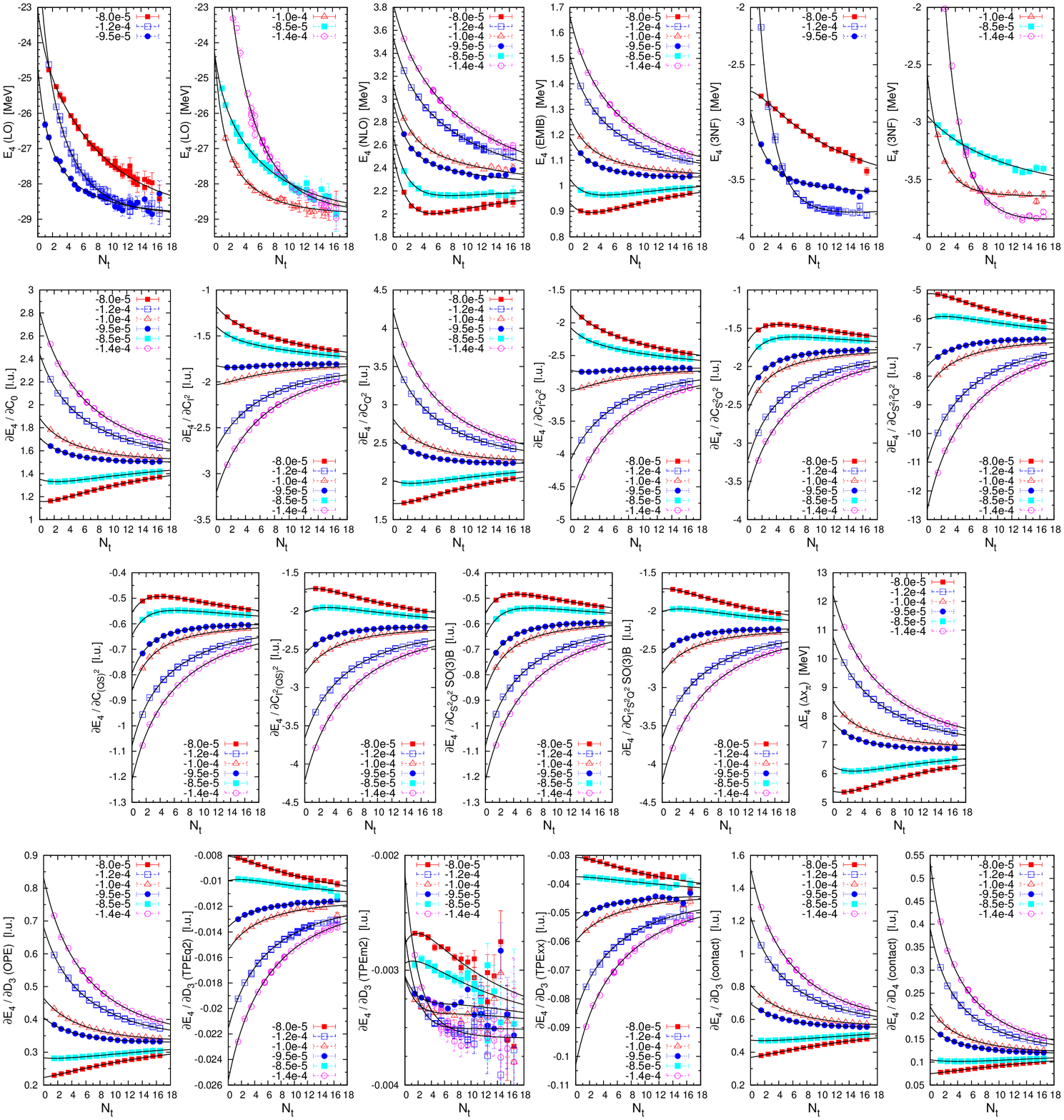}
\vspace{-.7cm}
\caption{Extrapolation of NLEFT results for $^{4}$He with $k_\mathrm{max}^{} = 3$. 
The definitions of the observables are given in the main text.
The LO energy is $E_\mathrm{LO}^{} = -28.87(6)$~MeV, and the result at NNLO including effective 4N interactions (as reported in Ref.~\cite{A28_letter}) is 
$E_\mathrm{NNLO+4N}^{} = -28.93(7)$~MeV. The empirical binding energy is $-28.30$~MeV. For each trial state, the value of $C_\mathrm{SU(4)}^{}$ is given in 
units of the (spatial) lattice spacing.
\label{4He}}
\end{figure}

\begin{figure}[t]
\includegraphics[width=\columnwidth]{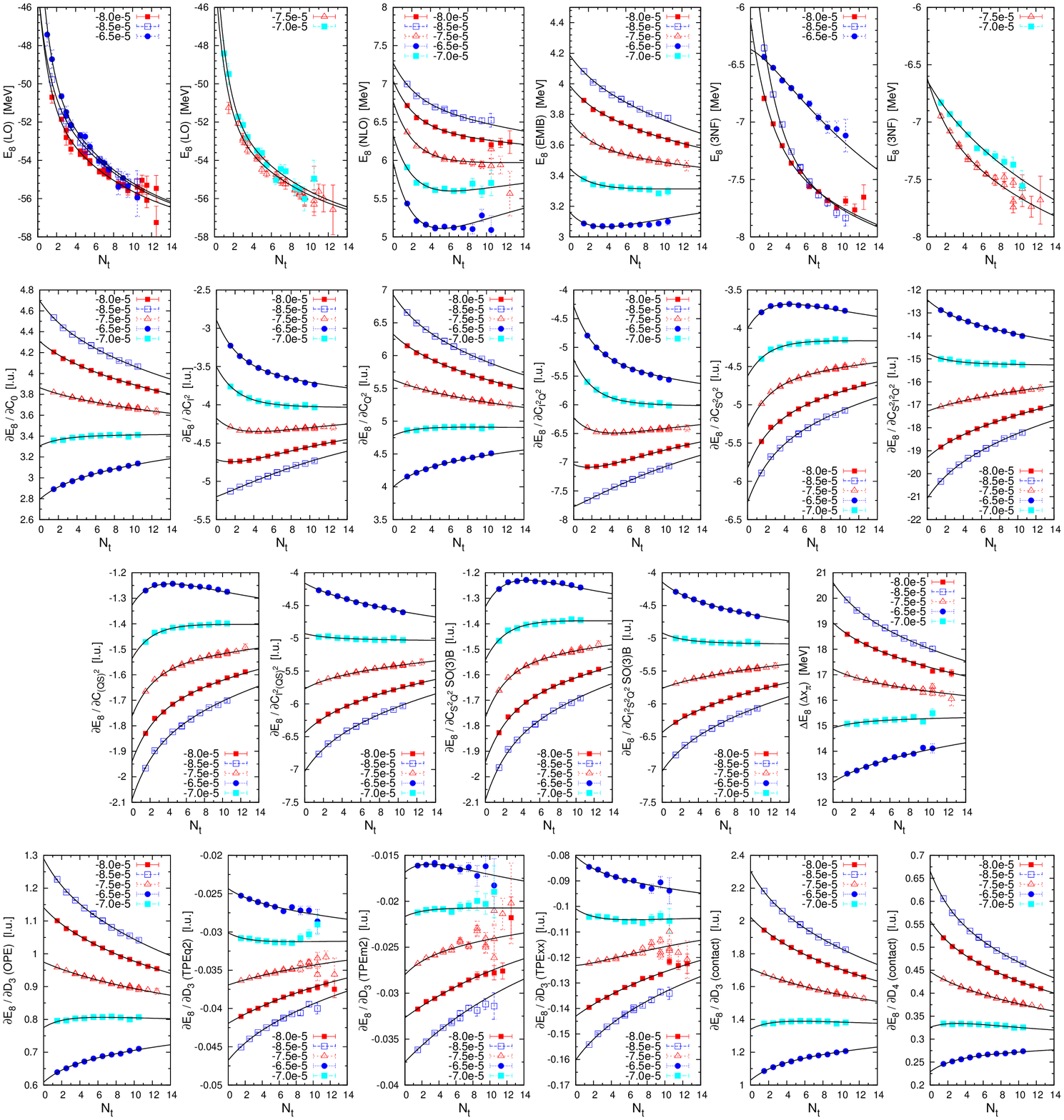}
\vspace{-.7cm}
\caption{Extrapolation of NLEFT results for $^{8}$Be with $k_\mathrm{max}^{} = 2$. 
The definitions of the observables are given in the main text.
The LO energy is $E_\mathrm{LO}^{} = -57.9(1)$~MeV, and the result at NNLO including effective 4N interactions (as reported in Ref.~\cite{A28_letter}) is
$E_\mathrm{NNLO+4N}^{} = -56.3(2)$~MeV. The empirical binding energy is
$-56.35$~MeV. For each trial state, the value of $C_\mathrm{SU(4)}^{}$ is given in units of the (spatial) lattice spacing.
\label{8Be}}
\end{figure}

\begin{figure}[t]
\includegraphics[width=\columnwidth]{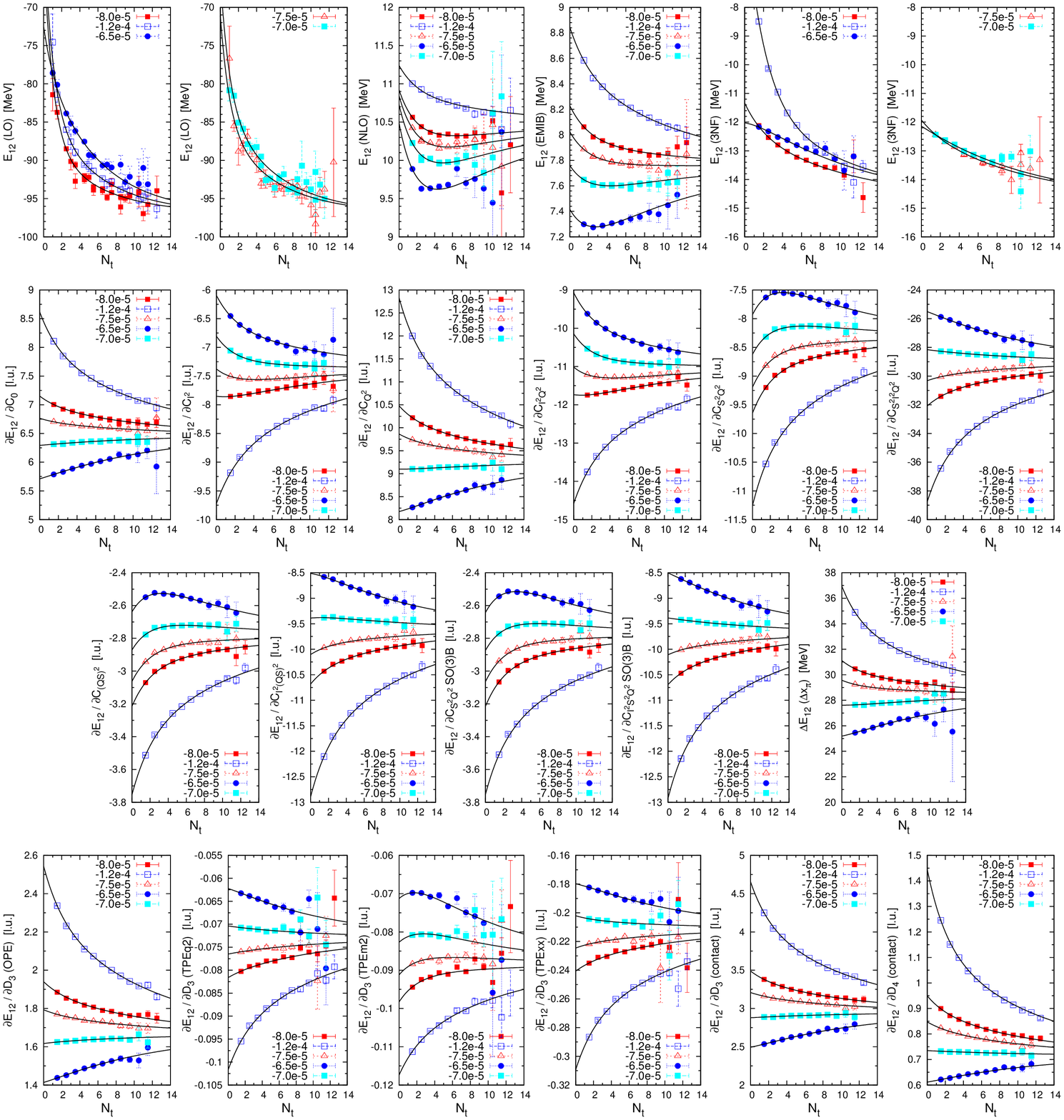}
\vspace{-.7cm}
\caption{Extrapolation of NLEFT results for $^{12}$C with $k_\mathrm{max}^{} = 2$. 
The definitions of the observables are given in the main text.
The LO energy is $E_\mathrm{LO}^{} = -96.9(2)$~MeV, and the result at NNLO including effective 4N interactions (as reported in Ref.~\cite{A28_letter}) is
$E_\mathrm{NNLO+4N}^{} = -90.3(2)$~MeV. The empirical binding energy is
$-92.16$~MeV. For each trial state, the value of $C_\mathrm{SU(4)}^{}$ is given in units of the (spatial) lattice spacing.
\label{12C}}
\end{figure}

\begin{figure}[t]
\includegraphics[width=\columnwidth]{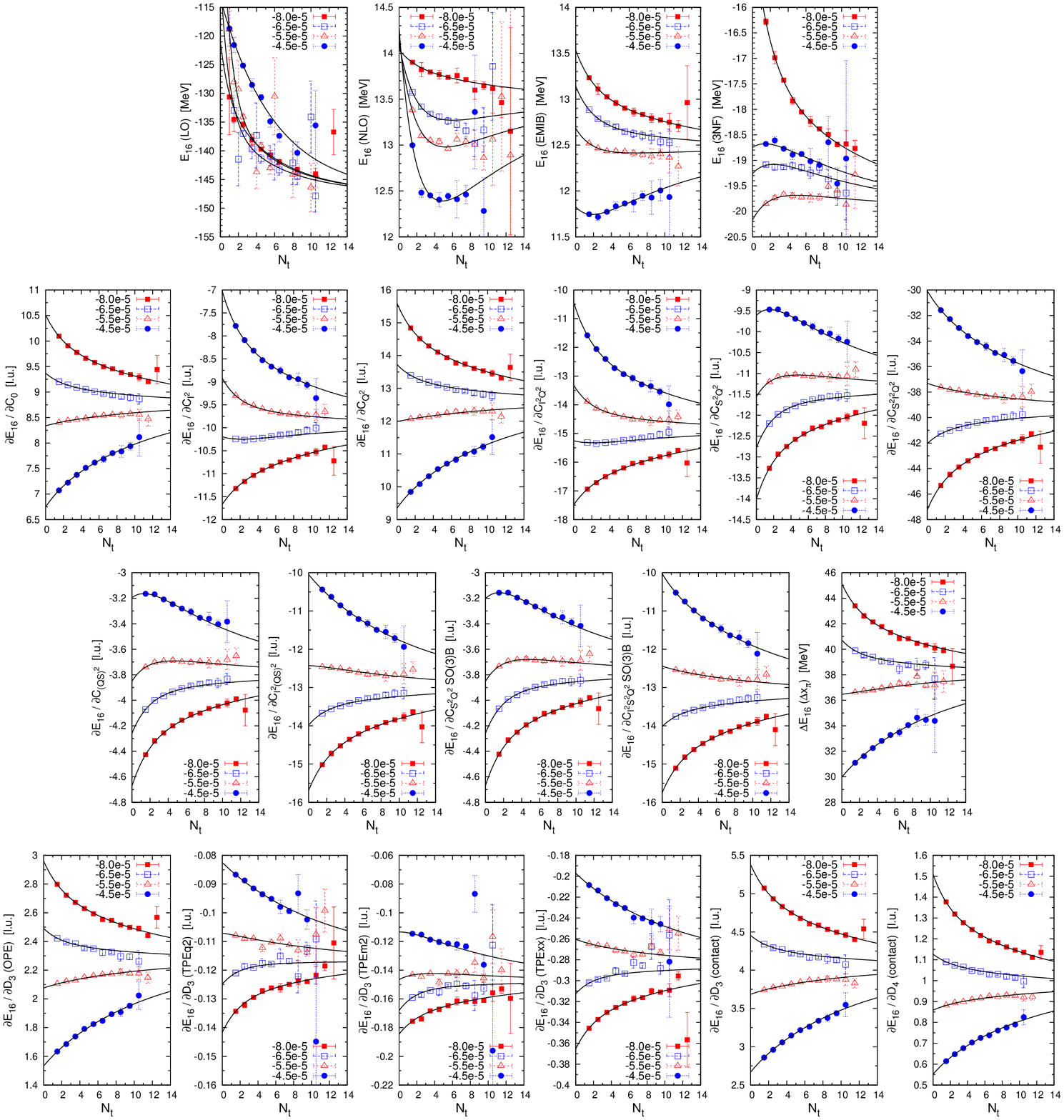}
\vspace{-.7cm}
\caption{Extrapolation of NLEFT results for $^{16}$O with $k_\mathrm{max}^{} = 2$.
The definitions of the observables are given in the main text.
The LO energy is $E_\mathrm{LO}^{} = -147.3(5)$~MeV, and the result
at NNLO including effective 4N interactions (as reported in Ref.~\cite{A28_letter}) is
$E_\mathrm{NNLO+4N}^{} = -131.3(5)$~MeV. The empirical binding energy is
$-127.62$~MeV. For each trial state, the value of $C_\mathrm{SU(4)}^{}$ is given in units of the (spatial) lattice spacing.
\label{16O}}
\end{figure}

Let us now consider the extrapolation formulas for the LO energy and the higher order perturbative corrections.
From Eqs.~(\ref{reps}) and~(\ref{delta}), we find for the LO energy
\begin{equation}
E_{A}^{j}(N_t^{}) = E_{A,0}^{} + \sum_{k=1}^{k_\mathrm{max}^{}}
|c_{A,j,k}^{}| \exp\left(-\frac{\Delta_{A,k}^{}N_t^{}}{\Lambda_t^{}}\right),
\label{extr_E}
\end{equation}
where $t = N_t^{} / \Lambda_t^{}$ with $\Lambda_t^{} = 150$~MeV, corresponding to $a_t^{}=1.32$~fm.
The energy gaps are defined as $\Delta_{A,k}^{} \equiv E_{A,k}^{} - E_{A,0}^{}$, and the index $j$ denotes
a specific choice of $t^\prime$ and $C_\mathrm{SU(4)}^{}$ in the trial wave function $|\Psi_A^{}(t^\prime)\rangle$. 
We take $\Delta_{A}^{k+1} > \Delta_{A}^{k}$, and $k_\mathrm{max}^{} = 3$ for $^4$He ($A = 4$) and $k_\mathrm{max}^{} = 2$ for $A \geq 8$. 
For the operator matrix elements that make up the perturbative NLO and NNLO corrections, we find
\begin{equation}
X_{A}^{{\mathcal{O}},j}(N_t^{}) = X_{A,0}^{\mathcal{O}} + \sum_{k=1}^{k_\mathrm{max}^{}}
x_{A,j,k}^{} \exp\left(-\frac{\Delta_{A,k}^{}N_t^{}}{2\Lambda_t^{}}\right),
\label{extr_X}
\end{equation}
where the dominant contributions are taken to be due to transition amplitudes involving the ground state and excited states.  In order for this to be a good approximation, it is 
necessary that the overlap between our trial state and the ground state not be small compared to the overlap with the low-lying excited states. 
It should be noted that the coefficients $x_{A,j,k}^{}$ can be positive as well as negative, which gives us the possibility of ``triangulating" the asymptotic values $X_{A,0}^{\mathcal{O}}$
from above and below. For this purpose, the parameters $t^\prime = N_t^\prime / \Lambda_t^{}$ and $C_\mathrm{SU(4)}^{}$ should be optimally
chosen for each value of $A$. For Eq.~(\ref{extr_E}), the dependence on $t$ is monotonically decreasing and thus no triangulation of the asymptotic value from above and 
below is possible. However, we are helped by the fact that the rate of convergence is twice that of Eq.~(\ref{extr_X}). In order to determine $E_{A,0}^{}$ and $X_{A,0}^{\mathcal{O}}$, a 
correlated $\chi^2$ fit to the LO energy and all NLO and NNLO matrix elements in NLEFT is performed for each value of $A$. This procedure also
determines the coefficients $c_{A,j,k}^{}, x_{A,j,k}^{}$ and $\Delta_{A,k}^{}$. 
We find that using $2$ to $6$ distinct trial states for each $A$ allows for a significantly more accurate and stable
determination of $E_{A,0}^{}$ and $X_{A,0}^{\mathcal{O}}$ than would be possible with a single trial state. 
Note that the energy gaps $\Delta_{A,k}^{}$ in the extrapolation functions are taken to be 
independent of the trial wave function $j$, which gives an additional consistency criterion. We find that a simultaneous description using Eqs.~(\ref{extr_E})
and~(\ref{extr_X}) accounts for all of the PMC data we have obtained for different $|\Psi_A^{}(t^\prime)\rangle$.


\section{Analysis of Projection Monte Carlo data \label{MC}}

\begin{figure}[t]
\begin{center}
\includegraphics[width=\columnwidth]{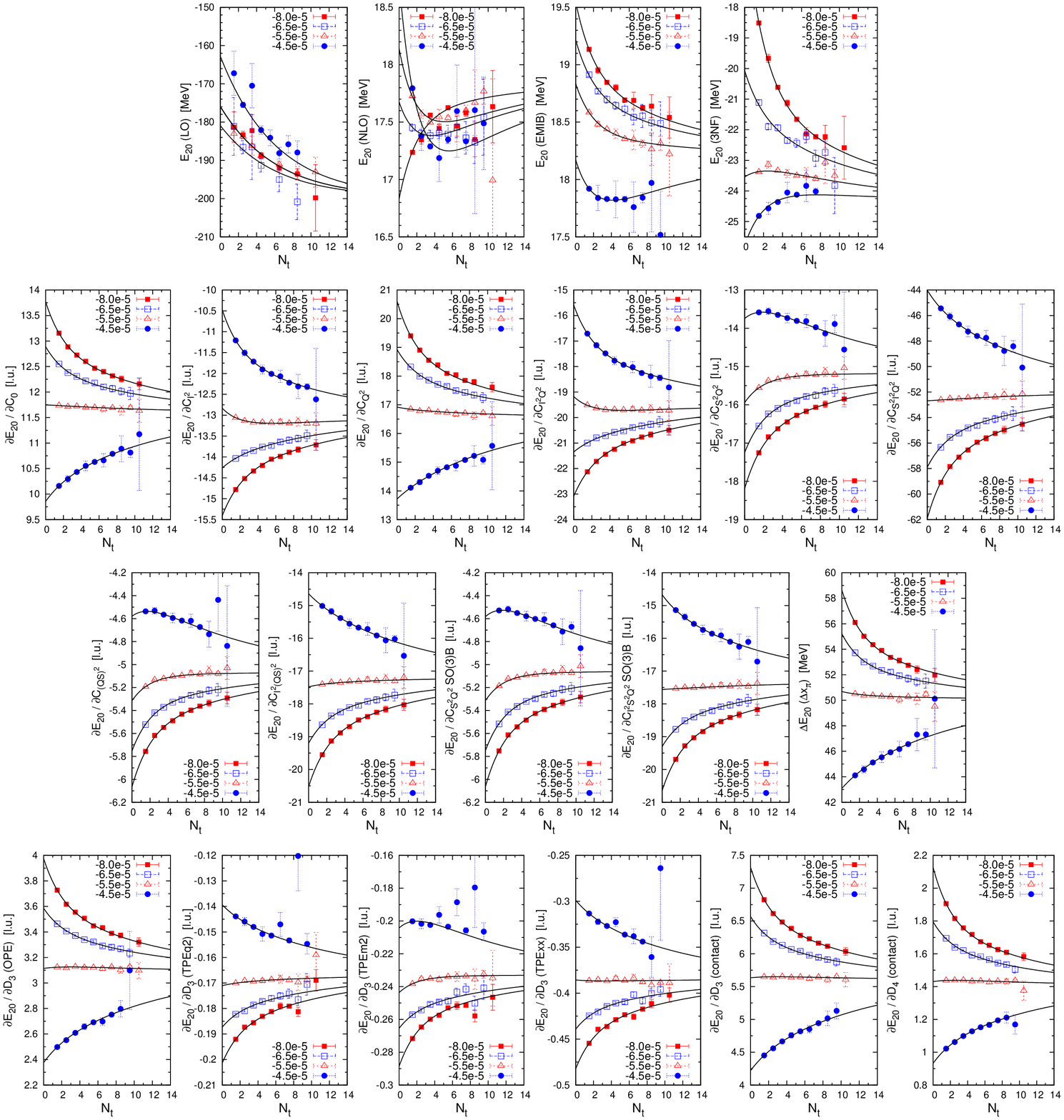}
\vspace{-.7cm}
\end{center}
\caption{Extrapolation of NLEFT results for $^{20}$Ne with $k_\mathrm{max}^{} = 2$.
The definitions of the observables are given in the main text.
The LO energy is $E_\mathrm{LO}^{} = -199.7(9)$~MeV, and the result
at NNLO including effective 4N interactions (as reported in Ref.~\cite{A28_letter}) is $E_\mathrm{NNLO+4N}^{} = -165.9(9)$~MeV. The empirical binding energy is
$-160.64$~MeV. For each trial state, the value of $C_\mathrm{SU(4)}^{}$ is given in units of the (spatial) lattice spacing.
\label{20Ne}}
\end{figure}

\begin{figure}[t]
\begin{center}
\includegraphics[width=\columnwidth]{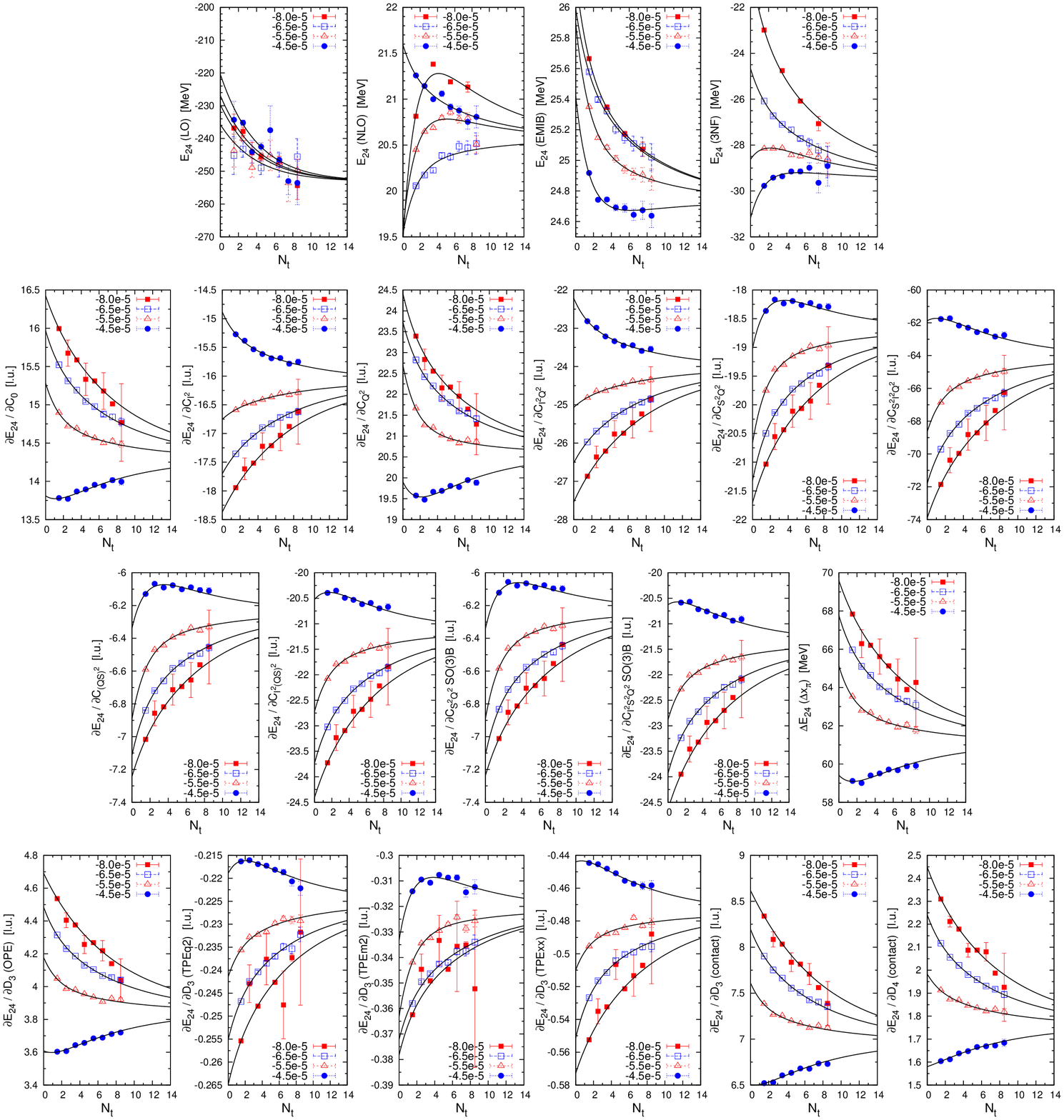}
\vspace{-.7cm}
\end{center}
\caption{Extrapolation of NLEFT results for $^{24}$Mg with $k_\mathrm{max}^{} = 2$.
The definitions of the observables are given in the main text.
The LO energy is $E_\mathrm{LO}^{} = -253(2)$~MeV, and the result
at NNLO including effective 4N interactions (as reported in Ref.~\cite{A28_letter}) is $E_\mathrm{NNLO+4N}^{} = -198(2)$~MeV. The empirical binding energy is
$-198.26$~MeV. For each trial state, the value of $C_\mathrm{SU(4)}^{}$ is given in units of the (spatial) lattice spacing.
\label{24Mg}}
\end{figure}

\begin{figure}[t]
\begin{center}
\includegraphics[width=\columnwidth]{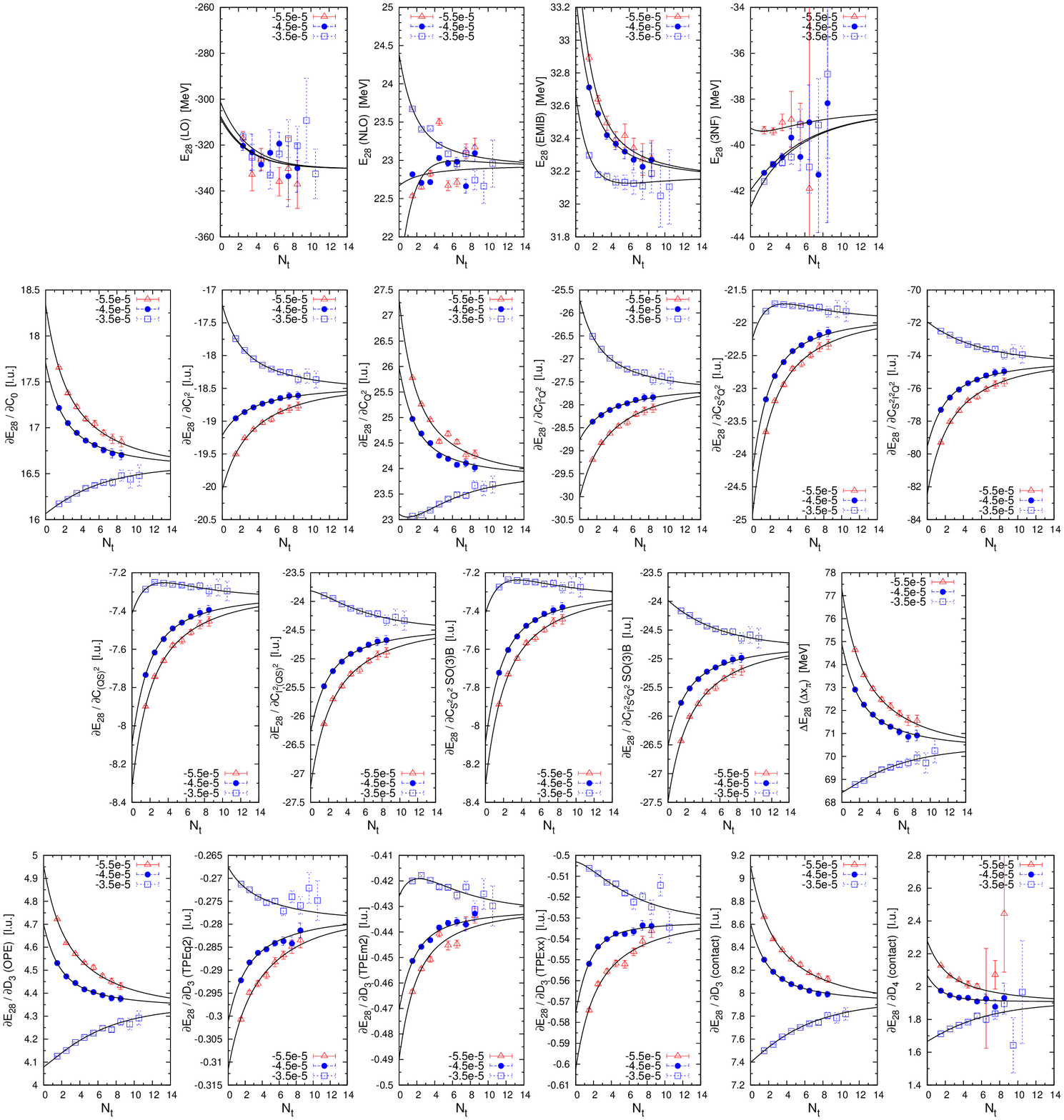}
\vspace{-.7cm}
\end{center}
\caption{Extrapolation of NLEFT results for $^{28}$Si with $k_\mathrm{max}^{} = 2$.
The definitions of the observables are given in the main text.
The LO energy is $E_\mathrm{LO}^{} = -330(3)$~MeV, and the result
at NNLO including effective 4N interactions (as reported in Ref.~\cite{A28_letter}) is $E_\mathrm{NNLO+4N}^{} = -233(3)$~MeV. The empirical binding energy is
$-236.54$~MeV. For each trial state, the value of $C_\mathrm{SU(4)}^{}$ is given in units of the (spatial) lattice spacing.
\label{28Si}}
\end{figure}

We shall next elaborate on how the extrapolation methods of Section~\ref{extra} perform when confronted with actual PMC data. First, we
show our data for the light nuclei $^{4}$He and $^{8}$Be in Figs.~\ref{4He} and~\ref{8Be}, respectively. Further, our results for 
$^{12}$C are given in Fig.~\ref{12C}, for $^{16}$O in Fig.~\ref{16O}, for $^{20}$Ne in Fig.~\ref{20Ne}, 
for $^{24}$Mg in Fig.~\ref{24Mg}, and for $^{28}$Si in Fig.~\ref{28Si}. The curves show a correlated $\chi^2$ fit for all trial states with a given $A$, using the same 
spectral density $\rho_A^{}(E)$. The upper row in each figure shows the LO energy, the total isospin-symmetric 
2NF correction (NLO), the electromagnetic and isospin-breaking corrections (EMIB) and the total 3NF correction. 
The remaining panels show the matrix elements $X_A^{\mathcal{O}}(t)$ that form part of the NLO and 3NF terms. 

In the second and third rows of Fig.~\ref{4He} through Fig.~\ref{28Si}, the operators $\partial E_{A}^{}/\partial C_i^{}$ give the contributions of the NLO contact interactions. The 
interactions that involve $C_0^{}$, $C_{I^2}^{}$, $C_{Q^2}^{}$, $C_{I^2Q^2}^{}$, $C_{S^2Q^2}^{}$, $C_{S^2I^2Q^2}^{}$, $C_{(QS)^2}^{}$ and 
$C_{I^2(QS)^2}^{}$ are defined in Eqs.~(18) to~(23) of Ref.~\cite{Epelbaum:2010xt}. Similarly, the interactions involving
$C_{S^2Q^2\rm{SO(3)B}}^{}$ and $C_{I^2S^2Q^2\rm{SO(3)B}}^{}$ are given in Eqs.~(55) and~(56) of Ref.~\cite{Epelbaum:2010xt}, and 
$\Delta E_{A}^{} (\Delta x_\pi^{})$ denotes the energy shift due the $\mathcal{O}(a^2)$-improved pion-nucleon coupling in Eq.~(57) of Ref.~\cite{Epelbaum:2010xt}. 
The operators $\partial E_{A}^{}/\partial D_i^{}$ give the individual contributions to the 3NF correction, which are defined in 
Eqs.~(37) to~(41) of Ref.~\cite{Epelbaum:2010xt}. 

The error estimates given in parentheses in the captions of Fig.~\ref{4He} through Fig.~\ref{28Si}
have been obtained from a $\chi^2$ minimization using the PMC errors for each datapoint as weight factors.
This procedure could be affected by autocorrelations in Monte Carlo time, and secondly by the fact that
the observables (except for the LO energy) are formed out of a ratio of amplitudes for which the PMC error is known separately, according to Eq.~(\ref{XAt}). 
Instead of simply adding these PMC errors in quadrature, a more realistic error estimate could be obtained by means of a resampling algorithm such as the Jackknife or the Bootstrap 
(for a pedagogical introduction, see {\it e.g.}~Ref.~\cite{stat}). In order to decrease the effects of any residual autocorrelations in the PMC data, the Jackknife
method can be combined with ``blocking'' of the data, whereby adjacent (in Monte Carlo time) samples are combined into blocks of increasing size until the
variance of the sample converges as a function of the block size (see {\it e.g.}~Ref.~\cite{DeTar}). 

In our PMC production runs, we do not use $k_\mathrm{max}^{} = 1$ (corresponding to a single energy gap) as this would in most cases lead to large values of $\chi^2$ as well
as inconsistent results for different choices of trial states (see Table~\ref{tab_12C}).
Extrapolations with $k_\mathrm{max}^{} = 2$ account very well for the medium-mass nuclei, which appear to be highly compact objects for which
the contamination from low-lying excited states is small. For $^4$He, we find that $k_\mathrm{max}^{} = 3$ is required in order to account for all the PMC data
for all observables and all trial states. However, while extrapolations with $k_\mathrm{max}^{} = 3$ may provide a more accurate description over a larger 
range of trial states, such fits are also much more difficult to constrain adequately, due to the much larger number of adjustable parameters involved. The extent and resolution
of our PMC data in Euclidean time also limits, in most cases, the number of resolvable energy gaps to $k_\mathrm{max}^{} = 2$.


\section{Statistical and systematic errors \label{errors}}

Our extrapolation procedure is examined in detail for the case of $^{12}$C in Table~\ref{tab_12C}. Our main fit, labeled ``fit 5'', is also shown in Fig.~\ref{12C}, and consists of
a simultaneous fit to five trial states that differ in the value of $C_\mathrm{SU(4)}^{}$. We observe that $\chi^2 \simeq 0.68$, which indicates that the error bars of
the individual MC data points are likely to be overestimated. This could plausibly happen as the observables are formed from the ratio of two amplitudes, the errors of which
are at present simply added in quadrature. A full-fledged jackknife error analysis may yield a more realistic result. The uncertainties shown in parentheses correspond to the
variances reported by the $\chi^2$ minimization. In order to assess the accuracy of these error estimates as well as the stability of the central values, we have generated a number 
of fits where each one of the five trial states has been excluded in turn, in the spirit of the Jackknife method~\cite{stat}.
These fits are denoted ``4a'' through ``4e'' in Table~\ref{tab_12C}. Evidently, these agree
closely with the full analysis, the largest discrepancy being due to the exclusion of trial state ``2'', which generates the largest shift in the extrapolated values as well as a
significant reduction in $\chi^2$. As is evident from Fig.~\ref{12C}, trial state ``2'' (denoted by blue open squares)
is furthest away from the ``triangulation point'', and may therefore not be completely described
by an extrapolation with $k_\mathrm{max}^{} = 2$. An even more stable result might be obtained by replacing that trial state in the analysis with one which is closer to the
triangulation point.

\begin{table}[t]
\caption{Uncertainty analysis of the Euclidean time extrapolation for $^{12}$C with $k_\mathrm{max}^{} = 2$.
The values of $C_\mathrm{SU(4)}^{}$ (in MeV$^{-2}$) for each trial state shown in Fig.~\ref{12C} are 
``1'' = $-8.0 \times 10^5_{}$,
``2'' = $-1.2 \times 10^4_{}$,
``9'' = $-7.5 \times 10^5_{}$,
``10'' = $-6.5 \times 10^5_{}$, and ``11'' = $-7.0 \times 10^5_{}$. The quantities shown (in MeV) are: The LO non-perturbative 2NF result, 
followed by the perturbative higher-order and isospin-breaking corrections as described in the main text. The fit labeled ``5'' 
(shown in Fig.~\ref{12C}) is a correlated extrapolation using
all trial states. The fits labeled ``4a'' - ``4e'' check the consistency of fit ``5'' under the removal of a single trial state from the full analysis.
We also show the (poorly constrained) extrapolations ``1a'' - ``1e'' where each trial state is treated separately. Note that this only allows
for an analysis with $k_\mathrm{max}^{} = 1$. The one-standard deviation error estimates (given in parentheses) are obtained from a 
Marquardt-Levenberg minimization of $\chi^2$ (per d.o.f.) with the Monte Carlo error estimates used as weights.
\label{tab_12C}}
\begin{center}
\begin{tabular}{c | c c c c c c}
Fit & Trial states & \multicolumn{1}{c}{LO (2NF)} & \multicolumn{1}{c}{NLO (2NF)} & \multicolumn{1}{c}{EMIB (2NF)} & 
\multicolumn{1}{c}{NNLO (3NF)} & $\chi^2$ \\ 
\hline\hline
$5$ & $1,2,9,10,11$ & $-96.9(2)$ & $10.48(3)$ & $7.76(1)$ & $-14.80(6)$ & 0.68 \\ \hline
$4a$ & $1,2,9,10$ & $-96.8(2)$ & $10.46(4)$ & $7.76(1)$ & $-14.84(6)$ & 0.71 \\
$4b$ & $1,2,9,11$ & $-97.0(2)$ & $10.45(3)$ & $7.76(1)$ & $-14.85(6)$ & 0.72 \\
$4c$ & $1,2,10,11$ & $-96.8(2)$ & $10.46(4)$ & $7.76(2)$ & $-14.80(6)$ & 0.72 \\
$4d$ & $1,9,10,11$ & $-97.3(2)$ & $10.54(5)$ & $7.72(2)$ & $-14.61(6)$ & 0.57 \\
$4e$ & $2,9,10,11$ & $-96.9(2)$ & $10.44(4)$ & $7.75(2)$ & $-14.94(7)$ & 0.69 \\ \hline
$1a$ & $1$ & $-95.0(2)$ & $10.17(2)$ & $7.79(1)$ & $-13.93(4)$ & 1.83 \\
$1b$ & $2$ & $-94.4(2)$ & $10.55(2)$ & $7.98(1)$ & $-14.46(5)$ & 3.35 \\
$1c$ & $9$ & $-93.5(2)$ & $10.03(2)$ & $7.72(1)$ & $-13.60(4)$ & 1.14 \\
$1d$ & $10$ & $-94.0(9)$ & $9.10(9)$ & $7.32(3)$ & $-13.73(16)$ & 2.36 \\
$1e$ & $11$ & $-92.1(2)$ & $9.83(2)$ & $7.58(1)$ & $-13.34(4)$ & 1.28
\end{tabular}
\end{center}
\end{table}

\begin{figure}[t!]
\begin{center}
\includegraphics[width=.85\columnwidth]{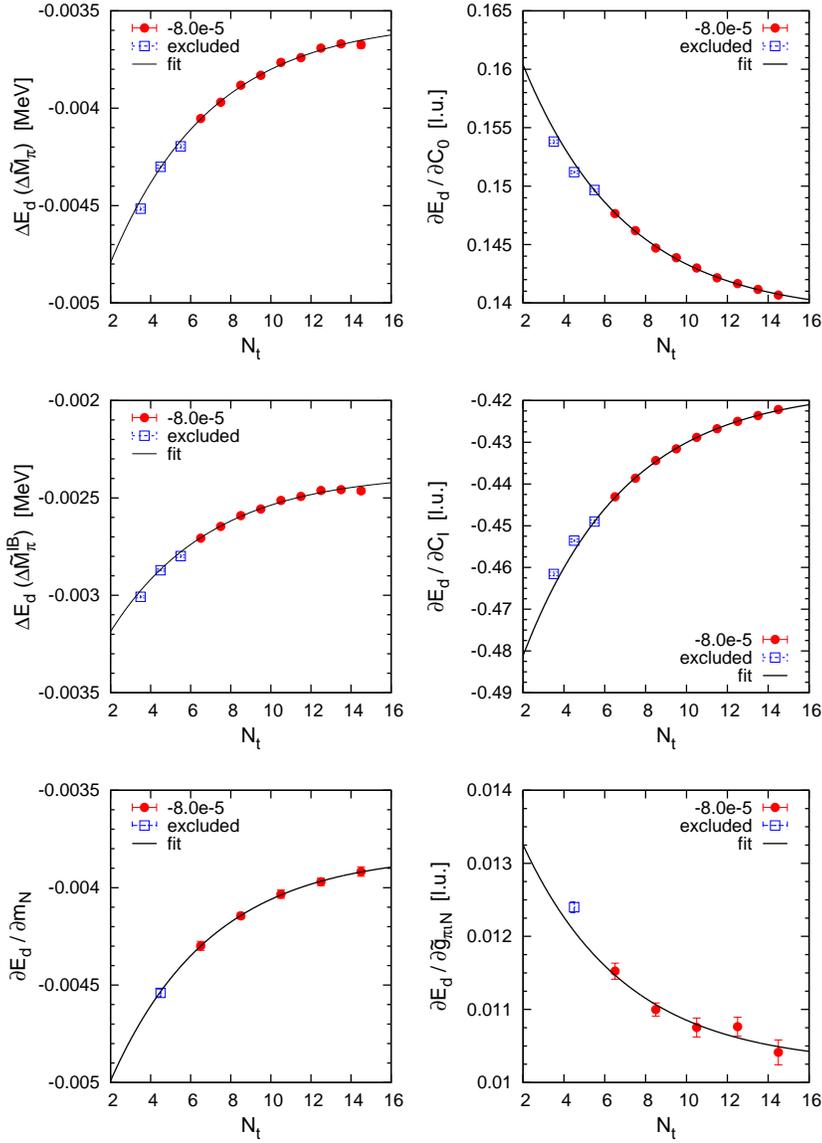}
\vspace{-.7cm}
\end{center}
\caption{Euclidean time extrapolation for the deuteron with $k_\mathrm{max}^{} = 1$ in a periodic $L = 3$ cube
with lattice spacing $a = 1.97$~fm, for the quantities given in Table~\ref{tab_deut}. The PMC data points at small Euclidean times (indicated by open squares) have been
excluded from the analysis in order to increase the stability of the results. The PMC data correspond to trial state ``1'' with $C_\mathrm{SU(4)}^{} = -8.0 \times 10^5_{}$~MeV$^{-2}$.
For full details and definitions, see Ref.~\cite{Epelbaum:2012iu}.
\label{deut}}
\end{figure}

\begin{table}[t]
\centering{
\caption{Euclidean time extrapolation for the deuteron with $k_\mathrm{max}^{} = 1$ in a periodic $L = 3$ cube
with lattice spacing $a = 1.97$~fm, with $\tilde g_{\pi N}^{} \equiv g_A^{}/(2f_\pi^{})$ and $m_N^{}$ the nucleon mass.
$E_d^{} (\mathrm{LO})$ denotes the (non-perturbative) LO energy, and the remaining quantities are perturbative
contributions which quantify the sensitivity of the of $E_d^{}$ to small shifts in the pion mass (for full details and definitions, see Ref.~\cite{Epelbaum:2012iu}).
The appropriate units are given for each quantity, with ``[l.u.]'' indicating units of the inverse (spatial) lattice spacing.
The second column shows the extrapolated Monte Carlo results, with one-standard-deviation errors similar to those in Table~\ref{tab_12C} given in parentheses.
The third column shows the results obtained from a Lanczos diagonalization of the two-nucleon Hamiltonian. 
\label{tab_deut}}
\vspace{.3cm}
\begin{tabular}{c | r | r}
Observable
& \multicolumn{1}{c|}{$^2$H (PMC+ex)}
& \multicolumn{1}{c}{$^2$H (Lanczos)} 
\\ \hline
$E_d^{} (\mathrm{LO})$ [MeV]
& $-9.070(12)$ & $-9.078$ \\
$\Delta E_d^{} (\Delta \tilde M_\pi^{})$ [MeV]
& $-0.003548(12)$ & $-0.003569$ \\  
$\Delta E_d^{} (\Delta \tilde M_\pi^\mathrm{IB})$ [MeV]
& $-0.002372(8)$ & $-0.002379$ \\ 
$\partial E_d^{} / \partial m_N^{}$
& $-0.00382(2)$ & $-0.003809$ \\
$\partial E_d^{} / \partial \tilde g_{\pi N}^{}$ [l.u.]
& $0.01024(11)$ & $0.01017$ \\
$\partial E_d^{} / \partial C_0^{}$ [l.u.]
& $0.13897(15)$ & $0.138867$ \\
$\partial E_d^{} / \partial C_I^{}$ [l.u.] 
& $-0.4171(4)$ & $-0.41660$\\
\end{tabular}
}
\end{table}

For comparison, we also show in Table~\ref{tab_12C} the results of independent, uncorrelated fits to each of the five trial states for $^{12}$C. These extrapolations do not benefit
from the consistency requirements of the multi-trial state extrapolations, and furthermore these can only be taken to $k_\mathrm{max}^{} = 1$, as the extent of the data in Euclidean time
is too short to constrain more than one energy gap. We observe that such extrapolations are clearly much less reliable, and suffer from several pronounced issues. One is the clear
tendency for ``spurious early convergence'', which is due to the lack of enforced independence on the value of $C_\mathrm{SU(4)}^{}$. We also observe that the extrapolated values
as well as the $\chi^2$ fluctuate significantly between extrapolations to different trial states. Again, trial state ``2'' appears to be the most problematic, although we also find that fits
with a smaller $\chi^2$ are no more reliable than those with a larger value. Neither do the variances produced by the $\chi^2$ minimization properly describe the uncertainties.

Also, as shown in Table~\ref{tab_12C}, the total uncertainty is dominated by that of the LO contribution, 
which does not consist of a ratio of amplitudes according to Eq.~(\ref{XAt}). This relatively large extrapolation error is due to the appearance of the
absolute values of the coefficients $c_{A,j,k}^{}$ in Eq.~(\ref{extr_E}), which prevents a ÒtriangulationÓ of the LO contribution. Nevertheless, the Jackknife method could be
used to evaluate the effect of autocorrelations between consecutive auxiliary-field configurations in Monte Carlo time on the LO result. We find that the 
elimination of such autocorrelations from the start, by allowing for sufficient decorrelation time between consecutive measurements, does not present any difficulties. For the NLO and NNLO 
operators, more consistent and reliable error estimates for the individual data points could clearly be obtained by Jackknife resampling of the Monte Carlo data. While such an analysis 
would not alter any of the conclusions concerning the stability and consistency of the Euclidean time extrapolations, we have investigated the likely outcome 
of a full Jackknife resampling of our data by performing multiple independent Monte Carlo runs for $^{12}$C with $N_t^{} = 12$. For such runs, we obtain LO energies for 
$N_t^{} = 11.0$, $11.5$ and $12.5$ by means of a numerical finite difference. Similarly, for the higher-order corrections we compute matrix elements for $N_t^{} = 11.5$ and $12.5$.
As expected, we find that the uncertainties of the LO energies are accurately given by the Monte Carlo errors, whereas those of the individual NLO matrix elements appear overestimated by
a factor of $\simeq 2$ due to cancellations between the numerator and denominator of Eq.~(\ref{XAt}). This result is consistent with the values of $\chi^2 < 1$ obtained from the
full extrapolation.

As the extent of our PMC data in Euclidean time is relatively short, we discuss next the expected reliability of our extrapolated results. Fortunately, the number 
of Euclidean time steps $N_t^{}$ available for the extrapolation does not decrease drastically with the number of nucleons $A$. At this time, our method has been
successfully applied to the spectrum, structure and electromagnetic properties of $^{12}$C in Refs.~\cite{Epelbaum:2012qn,Epelbaum:2012iu}, and to those of 
$^{16}$O in Ref.~\cite{16O_spectrum}, where consistency between delocalized plane-wave and alpha-cluster trial wave functions was established. 
In our ``triangulation'' method, the extrapolation is very strongly constrained by the requirement that all observables, for all trial states, should be described by the 
same exponential dependence on the Euclidean projection time $t$. Rapid convergence
in $t$ then translates into a small sensitivity to $C_\mathrm{SU(4)}^{}$ at large values of $t$, which helps to guard against ``spurious early convergence'', where
a smaller energy gap is overlooked. 

Nevertheless, in the absence of consistency conditions on the extrapolations, we find that our method is accurate even for 
$k_\mathrm{max}^{} = 1$ when the leading energy gap is very large, such as for the deuteron in a periodic $L = 3$ cube
with lattice spacing $a = 1.97$~fm (see Ref.~\cite{Epelbaum:2012iu}). In that case, the extrapolated results can be directly compared with Lanczos diagonalization, 
as shown in Table~\ref{tab_deut} and Fig.~\ref{deut}. In spite of this impressive agreement, we still need to 
consider the possibility that a sufficiently small energy gap can be missed in the extrapolation due to
the limited extent and resolution of the PMC data in Euclidean time. Of all the results presented here, those for $^8$Be are likely to
be most affected by the limited Euclidean projection time, as the convergence to the ground state is clearly the slowest and the data do not allow for
extrapolations beyond $k_\mathrm{max}^{} = 2$, unlike the case of $^4$He, where up to three energy gaps could be constrained by the PMC data.


\section{Summary and outlook \label{outlook}}

We have presented an overview of the techniques and the analysis used for Euclidean time extrapolations in NLEFT. 
The core issue is that, due to computational constraints, one must extract asymptotic values from a limited region in Euclidean projection time. 
In order to reduce the uncertainty of the extrapolation, we perform multi-exponential fits as prescribed by spectral decomposition for the asymptotic behavior of the 
projection amplitudes. We find that the fitted asymptotic values are greatly stabilized by using multiple initial states and observables. As examples of this analysis, we have 
shown energies at LO, NLO, and NNLO, as well as individual operator expectation values for the alpha nuclei up to $A=28$.

While these results are promising, the methods we have presented leave much room for further improvement. In particular, the current analysis is not adequate for the case 
when our trial states have only a small overlap with the ground state. In such cases, the raw PMC data will be far from their asymptotic values and it is unlikely that combining several 
different sets of data will provide any significant improvement. Another unfavorable situation that can arise is when the sign oscillations are severe and the quality of the PMC
data degrades very quickly with projection time. 

In these more difficult scenarios, one can improve on the situation by performing coupled multi-channel projections that evolve more than one initial state simultaneously 
in Euclidean time. This is the same approach which is being used in a technique called the ``adiabatic projection method'' used for describing scattering states of 
clusters~\cite{Rupak:2013aue,Pine:2013zja}. As shown in Ref.~\cite{Pine:2013zja}, when projecting $n$ states the exponential convergence of the ground state energy in projection 
time is given by the excitation energy of the $n^{\it{th}}$ excited state rather than by that of the first excited state. This has the potential to help significantly in circumventing the limited 
extent of the Euclidean time propagation.


\section*{Acknowledgments}

We are grateful for assistance in automated data collection by Thomas Luu.  We acknowledge partial financial support from the 
Deutsche Forschungsgemeinschaft (Sino-German CRC 110), the Helmholtz Association (Contract No.\ VH-VI-417), 
BMBF (Grant No.\ 05P12PDFTE), the U.S. Department of Energy (DE-FG02-03ER41260), and the U.S. National Science Foundation
(PHY-1307453). Further support
was provided by the EU HadronPhysics3 project and the ERC Project No.\ 259218 NUCLEAREFT. The computational resources 
were provided by the J\"{u}lich Supercomputing Centre at  Forschungszentrum J\"{u}lich and by RWTH Aachen.
T.~L. acknowledges a grant from the Magnus Ehrnrooth foundation of the Finnish Society of Sciences and Letters.


\end{document}